\documentclass[aps,prx,reprint,noeprint,floatfix]{revtex4-2}
\usepackage[utf8]{inputenc}
\usepackage{xcolor}
\usepackage{amsmath}
\usepackage{amssymb}
\usepackage{physics}
\usepackage{graphicx}
\usepackage{parskip}
\usepackage{hyperref}
\usepackage{varioref}       
\usepackage{cleveref}     

\crefname{table}{table}{tables}
\Crefname{table}{Table}{Tables}
\crefname{figure}{fig.}{figs.}
\Crefname{figure}{Fig.}{Figs.}
\crefname{section}{sect.}{sects.}
\Crefname{Section}{Sect.}{Sects.}

\definecolor{niceblue}{rgb}{0.016, 0.30, 0.56}
\definecolor{nicered}{rgb}{0.74, 0, 0.31}
\hypersetup{colorlinks = true, linkcolor = nicered, citecolor = niceblue, urlcolor = nicered, linktocpage = true}

\usepackage{mathptmx}

\newcommand{\commie}[1]{}
\newcommand{\ads}{\mathrm{AdS}}

\newenvironment{eqaed}
    {\begin{equation}
    \begin{aligned}
    }
    { 
    \end{aligned}
    \end{equation}
    \ignorespacesafterend
    }

\begin{document}

\title{\bf Emergent strings at infinite distance with broken supersymmetry}
\author{Ivano Basile}
\email{ivano.basile@umons.ac.be}
\affiliation{Service de Physique de l'Univers, Champs et Gravitation, Universit\'{e} de Mons - UMONS, Place du Parc 20, 7000 Mons, Belgium}

\begin{abstract}

    We investigate the infinite-distance properties of families of unstable flux vacua in string theory with broken supersymmetry. To this end, we employ a generalized notion of distance in the moduli space and we build a holographic description for the non-perturbative regime of the tunneling cascade in terms of a renormalization group flow. In one limit we recover an exponentially light tower of Kaluza-Klein states, while in the opposite limit we find a tower of higher-spin excitations of D1-branes, realizing the emergent string proposal. In particular, the holographic description includes a free sector, whose emergent superconformal symmetry resonates with supersymmetric stability, the CFT distance conjecture and S-duality. We compute the anomalous dimensions of scalar vertex operators and single-trace higher-spin currents, finding an exponential suppression with the distance which is not generic from the renormalization group perspective, but appears specific to our settings.

\end{abstract}

\maketitle

\section{Introduction}\label{sec:introduction}
    
The last decade of research in string theory, and quantum gravity in general, has witnessed a remarkable breadth of progress and novel ideas. A variety of connections between fundamental interactions of microscopic degrees of freedom and the breakdown of the corresponding low-energy effective field theory (EFT) dynamics have been proposed and thoroughly investigated. As a result, the existing newtork of swampland criteria~\cite{Vafa:2005ui}~\footnote{See~\cite{Palti:2019pca, vanBeest:2021lhn, Grana:2021zvf} for reviews.} to determine consistent EFTs has been expanded and enriched. In particular, numerous insights have been collected about infinite-distance asymptotic regions of moduli space in EFTs coupled to gravity~\cite{Klaewer:2016kiy,Blumenhagen:2018nts,Heidenreich:2018kpg,Grimm:2018cpv,Corvilain:2018lgw,Grimm:2018ohb, Grimm:2019wtx, Grimm:2019ixq, Gendler:2020dfp, Grimm:2020ouv, Bastian:2021eom,Klaewer:2020lfg,Cecotti:2020rjq,Klaewer:2021vkr,Lanza:2020qmt,Lanza:2020htb,Heidenreich:2021yda} in support of the distance conjecture~\cite{Ooguri:2006in} and its extensions~\cite{Baume:2016psm,Klaewer:2016kiy,Lee:2018urn, Lee:2019xtm, Lee:2019wij, Lust:2019zwm,Andriot:2020lea, Calderon-Infante:2020dhm, Baume:2020dqd, Perlmutter:2020buo}.
    
The emerging picture appears to be intimately tied to string dualities, and suggests that quantum-gravitational consistency entails a very specific breakdown of EFT. Namely, an infinite tower of massive states would become parametrically light at an exponential rate in the proper distance in moduli space, and furthermore the states would pertain either to a Kaluza-Klein (KK) tower or higher-spin excitations of tensionless strings~\footnote{Recently, a deeper understanding of tensionless limits of strings in $\ads$ has been achieved~\cite{Gaberdiel:2014cha, Gaberdiel:2015mra, Gaberdiel:2015uca, Ferreira:2017pgt, Gaberdiel:2018rqv, Eberhardt:2018ouy, Eberhardt:2019ywk, Eberhardt:2020bgq, Eberhardt:2021jvj, Gaberdiel:2021iil, Gaberdiel:2021jrv}.}. These towers of states signal, respectively, the presence of extra dimensions of space or extended objects in the spectrum.
    
Despite many advances, investigations have focused on supersymmetric settings. In order to achieve a deeper understanding, and ultimately connect these ideas with phenomenology, it is paramount to address supersymmetry breaking, which at present lacks a comprehensive guiding principle. Among the wide variety of mechanisms that have been proposed, string-scale supersymmetry breaking appears to provide a natural setting to seek instructive lessons beyond the current ``lamppost''~\cite{Montero:2020icj, Hamada:2021bbz, Tarazi:2021duw, Bedroya:2021fbu}~\footnote{To this end, a novel proposal based on domestic geometry has been put forth in~\cite{Cecotti:2021cvv}, and another finiteness criterion for the landscape has been recently suggested in~\cite{Grimm:2021vpn}.}. To wit, na\"{i}ve dimensional arguments have been recently supplemented by additional considerations~\cite{Cribiori:2021gbf, Castellano:2021yye, DallAgata:2021nnr} on the (in)consistency of light gravitini within low-energy supersymmetry breaking, and the gravitino mass appears to play an important role reminiscent of ``brane supersymmetry breaking'' (BSB)~\cite{Antoniadis:1999xk, Angelantonj:1999jh, Aldazabal:1999jr, Angelantonj:1999ms}, as discussed in~\cite{Coudarchet:2021qwc}.
    
Given the present state of affairs, we are compelled to attempt extending the investigation of swampland proposals to non-supersymmetric settings, and to this end the $\text{SO}(16) \times \text{SO}(16)$ heterotic model of~\cite{AlvarezGaume:1986jb, Dixon:1986iz}, the $\text{U}(32)$ ``type $0'\text{B}$'' model of~\cite{Sagnotti:1995ga, Sagnotti:1996qj} and the $\text{USp}(32)$ model of~\cite{Sugimoto:1999tx} stand as promising candidates~\footnote{See also~\cite{Mourad:2017rrl, Basile:2021vxh, Mourad:2021lma} for reviews.}. In particular, the latter features a simple realization of BSB, whereby the closed-string sector remains supersymmetric while supersymmetry is broken in the open-string sector. The appearance of a Goldstino singlet in the perturbative spectrum hints at a spontaneous breaking, and the low-energy physics features the expected interactions \textit{\`{a} la} Volkov-Akulov~\cite{Dudas:2000nv, Pradisi:2001yv}, but a satistactory description of the corresponding super-Higgs mechanism in ten-dimensions remains elusive~\cite{Dudas:2000ff, Dudas:2001wd}.
    
Various swampland conjectures have been studied in these models~\cite{Basile:2020mpt,Basile:2021mkd} and in other settings with supersymmetry breaking~\cite{Bonnefoy:2020fwt}. In particular, some hints regarding light towers of states have been discussed in~\cite{Basile:2020mpt}. In this paper we shall focus on infinite-distance limits. In this respect, one expects supersymmetry breaking to dramatically affect vacua~\cite{Fischler:1986ci, Fischler:1986tb, Dudas:2004nd, Kitazawa:2008hv, Pius:2014gza} and destroy exact moduli spaces. A milder counterpart of this scenario would involve potentials lifting the moduli, and their role has been discussed in~\cite{Calderon-Infante:2020dhm, Lanza:2020htb}. In order to circumvent these limitations, we shall investigate generalized notions of distance using holography. A similar proposal has been put forth in~\cite{Baume:2020dqd, Perlmutter:2020buo}~\footnote{Let us mention that a complementary approach based on bootstrap methods has also provided some insights~\cite{Guerrieri:2021ivu, Caron-Huot:2021rmr, Kundu:2021qpi, Caron-Huot:2021enk}.} using the Zamolodchikov metric~\cite{Zamolodchikov:1986gt}. This approach to the geometry of theory space has been extended by O'Connor and Stephens~\cite{oconnor} in the context of quantum information theory, and by Anselmi~\cite{Anselmi:1997rd, Anselmi:1999xk, Anselmi:2000fu, Anselmi:2011bp} in the context of renormalization group (RG) flows. The former metric has been recently revisited by Stout~\cite{Stout:2021ubb} (see also~\cite{Balasubramanian:2014bfa, Erdmenger:2020vmo}), and in this work we shall employ both metrics to explore infinite-distance limits in the absence of supersymmetry.
    
Our findings reveal that the breakdown of EFT at infinite distance involves either a KK tower arising from compact extra dimensions or higher-spin excitations of a D1-brane. The latter lies in a stringy regime that we approach holographically, and arises as the endpoint of a cascade of flux tunneling processes in unstable brane configurations driven by weak gravity~\cite{Antonelli:2019nar, Basile:2021mkd}. Remarkably, in the Sugimoto model of~\cite{Sugimoto:1999tx} supersymmetry is restored, thereby granting stability as expected from the considerations of~\cite{Ooguri:2016pdq}.
    
After discussing in detail our setup from the bulk and holographic perspectives, we compute the (generalized) distances associated to the endpoints of the tunneling cascade, finding that they diverge. Then, we compute anomalous dimensions of scalar operators and (single-trace) higher-spin currents in the dual field theory, and we entertain the possibility of a novel heterotic-orientifold S-duality in the final state. While our results hold in more general settings, the asymptotic scalings specific to the string models of interest \emph{exponentially} suppress the anomalous dimensions in a precise sense that we discuss. Furthermore, these novel realizations of the emergent string proposal~\cite{Lee:2018urn, Lee:2019xtm, Lee:2019wij}~\footnote{See also~\cite{Lee:2021qkx, Lee:2021usk, Alvarez-Garcia:2021pxo, Lee:2021uuu} for novel results in this direction in the context of F-theory.}, supersymmetric protection and heterotic-orientifold duality~\cite{Polchinski:1995df} exhibit a tantalizing interplay via $\text{Spin}(8)$ triality. 
    
\section{Brane dynamics and weak gravity}\label{sec:branes_wgc}

Our starting point is to investigate the vacua of the string models that we have introduced in the preceding section. In more familiar settings, these comprise the trivial configuration where spacetime is flat and all fields vanish. However, in the present models, the low-energy EFT contains the (Einstein-frame) dilaton potential~\cite{Dudas:2000nv, Pradisi:2001yv, Mourad:2017rrl, Basile:2021vxh, Mourad:2021lma}
\begin{eqaed}\label{eq:tadpole_potential}
    V_\text{eff}(\phi) = T \, e^{\gamma \phi} \, ,
\end{eqaed}
where the parameter $\gamma = \frac{3}{2} \, , \, \frac{5}{2}$ for the orientifold models and the heterotic model respectively. For the former, the coefficient $T = \mathcal{O}({\alpha'}^{-1})$ can be interpreted as the residual tension of the $\overline{\text{D}9}$-branes and the O9-plane, while for the latter it can be interpreted as the one-loop vacuum energy (density). This potential, absent a balancing act, drives the vacuum to a runaway, where $\phi \to - \infty$. Therefore, standard string perturbation theory is compromised by a \emph{dynamical tadpole}, whose dramatic gravitational backreaction is yet to be understood completely. In the pioneering work of~\cite{Dudas:2000ff}~\footnote{See also~\cite{Blumenhagen:2000dc} for a T-dual version of this construction.}, it was found that the most symmetric solutions appear to entail a spontaneous compactification of one spatial dimension into an interval, which however hosts curvature and/or coupling singularities at its endpoints~\footnote{The role of fermions in geometries of this type has been studied in~\cite{Mourad:2020cjq}.}. These solutions have been generalized to families~\cite{Antonelli:2019nar, Basile:2021vxh, Mourad:2021lma, Mourad:2021qwf, Mourad:2021roa}, strongly suggesting that geometries of these type are sourced by branes, whose presence breaks the isometry group accordingly. On the one hand, the resulting geometries feature a universal finite-distance ``pinch-off'' singularity~\cite{Antonelli:2019nar}, which dovetails nicely with the recent considerations of~\cite{Buratti:2021fiv, Buratti:2021yia} in the context of the swampland.

\begin{figure}[!ht]
\centering
\includegraphics[width=.4\textwidth]{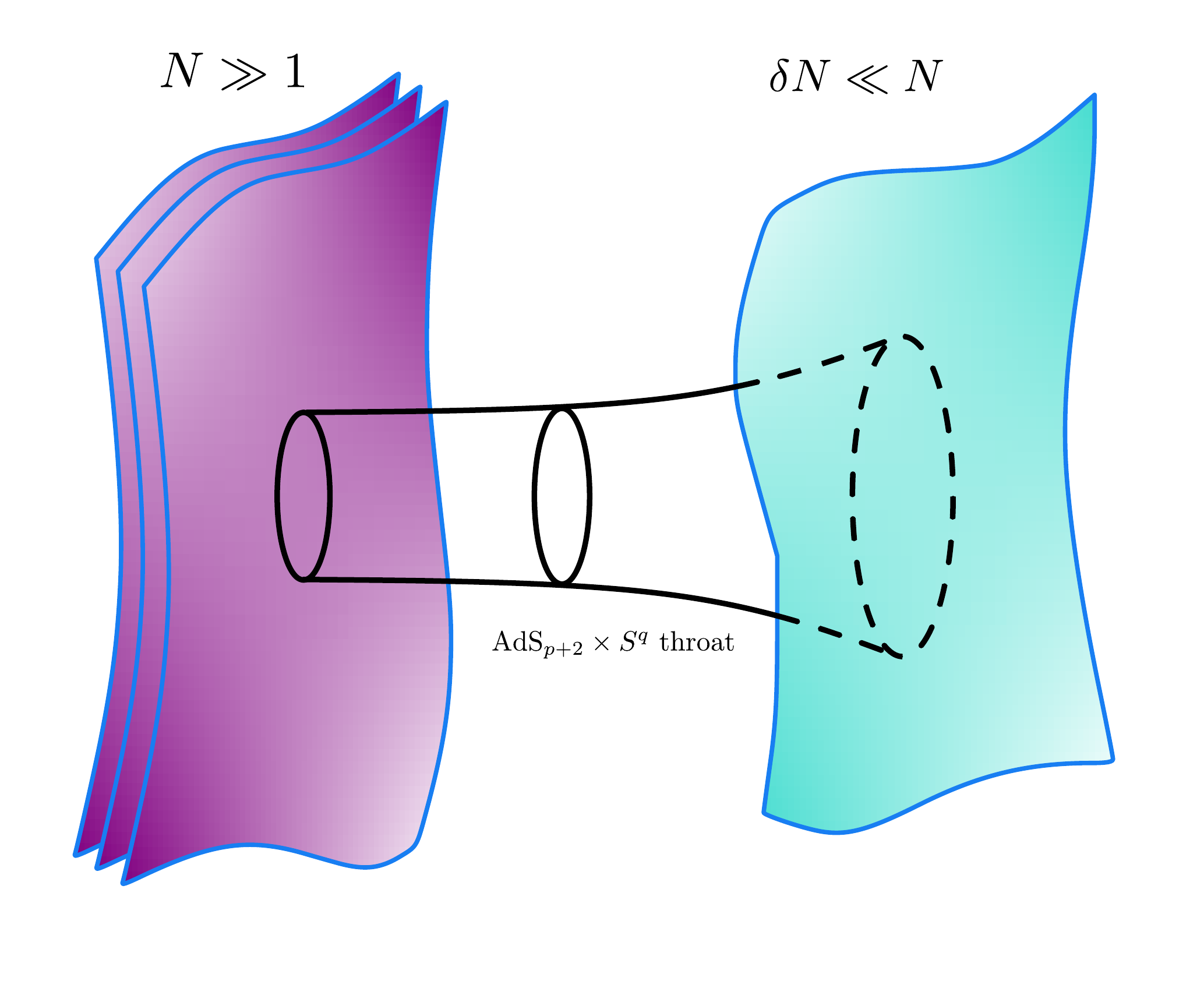}
\caption{A heavy stack of D1-branes (in the orientifold models) or NS5-branes (in the heterotic model) sources a spacetime geometry whose near-horizon limit is an $\ads \times S$ throat~\cite{Antonelli:2019nar, Basile:2021mkd}. One can expect branes on conical singularities to produce similar Freund-Rubin compactifications in this limit.}
\label{fig:throat_branes}
\end{figure}

\begin{figure}[!ht]
\centering
\includegraphics[width=.4\textwidth]{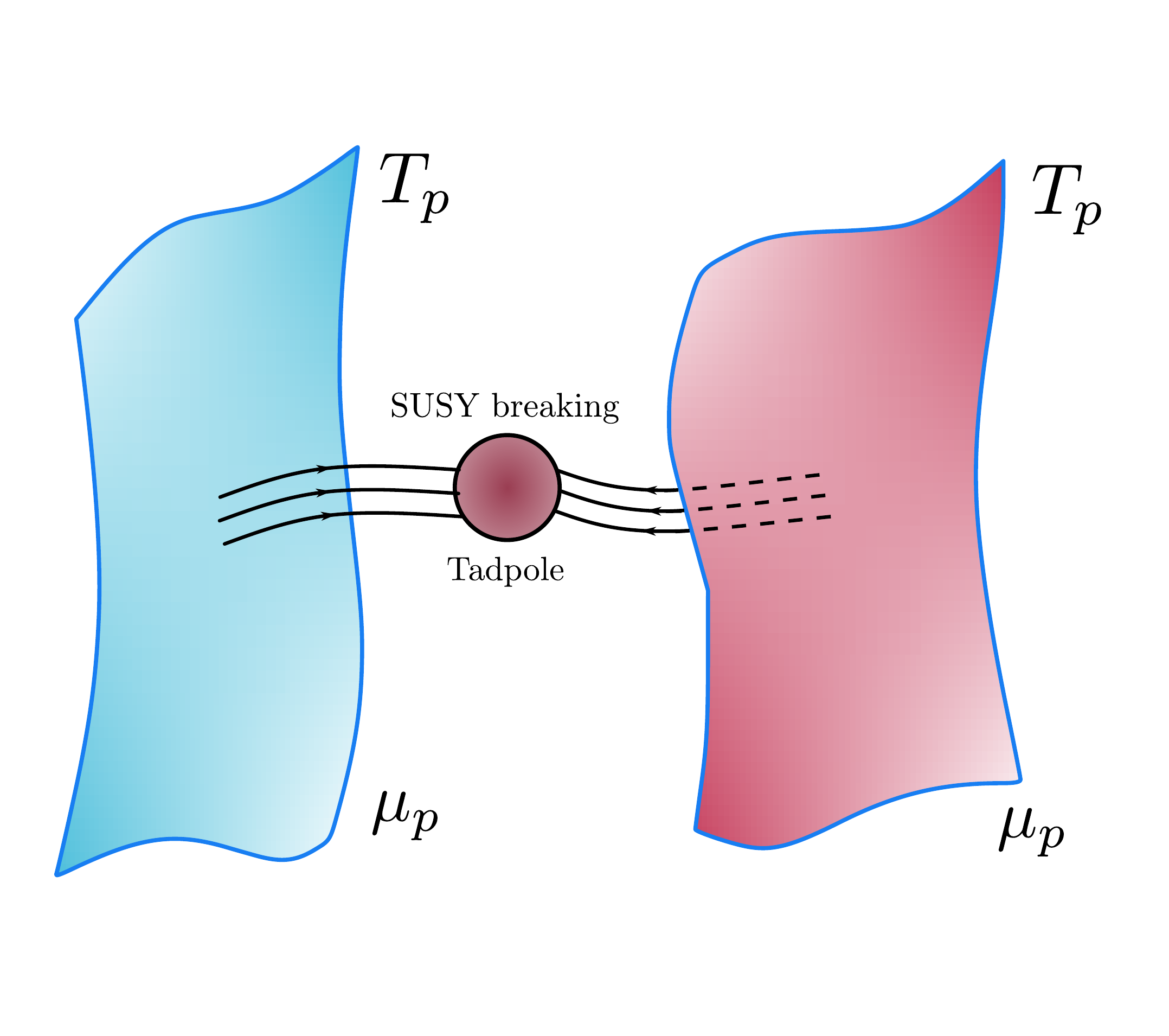}
\caption{The interaction between branes in the presence of string-scale supersymmetry breaking is mediated by the gravitational tadpole. As as a result, the effective charge-to-tension ratio is renormalized by a $\mathcal{O}(1)$ factor, and like-charge branes exert mutually repulsive forces~\cite{Antonelli:2019nar, Basile:2021mkd}.}
\label{fig:parallel_branes}
\end{figure}

On the other hand, the backreaction of extremal charged branes is somewhat milder, since the finite-distance ``pinch-off'' is accompanied by a near-horizon throat which is weakly curved and weakly coupled when the number of branes is large. The resulting geometry is depicted in \cref{fig:throat_branes}. The string models that we have discussed in the introduction contain a variety of such branes, whose presence can be ascertained from the consistency of their perturbative spectra~\cite{Dudas:2000sn, Dudas:2001wd} via orientifold techniques applied to one-loop vacuum amplitudes~\cite{Sagnotti:1987tw, Pradisi:1988xd, Horava:1989vt, Horava:1989ga, Bianchi:1990yu, Bianchi:1990tb, Bianchi:1991eu, Sagnotti:1992qw}. In particular, the \text{USp}(32) model of Sugimoto~\cite{Sugimoto:1999tx} contains charged D1-branes and D5-branes, while the $\text{U}(32)$ type $0'\text{B}$ model~\cite{Sagnotti:1995ga, Sagnotti:1996qj} also contains charged D3-branes and D7-branes. While D5-branes and D7-branes are more subtle in this respect, D3-branes source a quasi-$\ads_5 \times S^5$ near-horizon throat~\cite{Angelantonj:1999qg, Angelantonj:2000kh}, while D1-branes source a \emph{bona fide} $\ads_3 \times S^7$ throat~\cite{Mourad:2016xbk, Antonelli:2019nar, Basile:2021vxh, Basile:2021mkd}. Similarly, NS5-branes in the heterotic model source an $\ads_7 \times S^3$ throat~\footnote{Non-supersymmetric $\ads$ compactifications can be also found in supersymmetric models~\cite{Apruzzi:2019ecr, Apruzzi:2021nle}.}. All these geometries have no scale separation, and for a large number $N$ of branes, where the EFT regime is expected to be reliable, the string coupling $g_s$ and the radius $R$ of the internal $S^q$ scale according to~\cite{Mourad:2016xbk, Antonelli:2019nar, Basile:2021vxh, Basile:2021mkd}
\begin{eqaed}\label{eq:ads_s_orientifold}
    g_s & \propto N^{- \frac{2}{(q - 1) \gamma - 1}} \ll 1 \, , \\
    {\alpha'}^{- \frac{1}{2}} R & \propto N^{\frac{4 \gamma}{(q - 1) \gamma - 1}} \gg 1 \, .
\end{eqaed}
The resulting geometries are unstable, both perturbatively~\cite{Basile:2018irz} and non-perturbatively~\cite{Antonelli:2019nar}. In the former case, field fluctuations violating the Breitenlohner-Freedman bound~\cite{Breitenlohner:1982jf} can potentially be avoided replacing the internal $S^q$ with a suitable Einstein manifold $\mathcal{M}_q$~\footnote{Recent efforts in the bootstrap of Laplacian spectra~\cite{Bonifacio:2020xoc, Bonifacio:2021msa} could be fruitful in this respect. For hyperbolic manifolds, rigidity would simplify the analysis~\cite{Mostow:1968gg, DeLuca:2021pej}.}, which is expected to arise placing branes on a conical singularity~\cite{Klebanov:1998hh}, or performing a suitable orbifold projection~\footnote{In the heterotic model this is readily achieved by an antipodal $\mathbb{Z}_2$ projection, while the orientifold models are subtler. At any rate, double-trace operators could present additional complications~\cite{Witten:2001ua}. We thank Edward Witten for pointing this out.}. The latter case, however, cannot be avoided, and flux tunneling occurs with a probability $\Gamma$ per unit volume per unit time which is schematically of order~\cite{Coleman:1977py, Callan:1977pt, Coleman:1980aw, Brown:1987dd, Brown:1988kg, Antonelli:2019nar}
\begin{eqaed}\label{eq:decay_rate}
    \log \Gamma \overset{N \gg 1}{\sim} - \, N^{\frac{(p + 1) \gamma + 1}{(q - 1) \gamma - 1}}
\end{eqaed}
for $\ads_{p+2} \times \mathcal{M}_q$ in the semiclassical limit $N \gg 1$. After nucleation branes expand, and one can determine what forces they exert on each other computing static interaction potentials between parallel stacks~\cite{Antonelli:2019nar, Basile:2021mkd}. One finds that branes with the same charges repel, consistently with the weak gravity~\cite{Arkani-Hamed:2006emk} and repulsive force~\cite{Heidenreich:2019zkl} conjectures. The net repulsion is mediated by the supersymmetry-breaking dynamical tadpole, which renormalizes the effective charge-to-tension ratio by the $\mathcal{O}(1)$ factor
\begin{eqaed}\label{eq:charge-to-tension_ratio}
    \left(\frac{\mu_p}{T_p}\right)_\text{eff} = \sqrt{\frac{16 \gamma}{(p+1) ((q - 1) \gamma - 1)}} \left(\frac{\mu_p}{T_p}\right)_\text{bare} \, ,
\end{eqaed}
as shown in \cref{fig:parallel_branes}.

\section{Bubble nucleation and holography}\label{sec:bubbles_holography}

According to our preceding discussion, the $\ads$ vacua at stake undergo flux tunneling, nucleating charged branes. Although the corresponding decay rates are parametrically exponentially suppressed~\footnote{In order to conclude that the vacua are parametrically long-lived, one would need to exclude charged bubbles of nothing along the lines of~\cite{Horowitz:2007pr, Bomans:2021ara}.}, eventually this non-perturbative instability drives the vacua to progressively lower values of $N$, at least until the low-energy EFT ceases to be reliable. This process points to a dynamically generated trajectory in the (discrete) landscape of flux vacua, somewhat in the spirit of~\cite{Lanza:2020qmt, Lanza:2020htb}. One is thus naturally led to investigate whether the endpoints of this trajectory, located at $N \to \infty$ and $N=1$, lie at infinite distance in some sense, and whether towers of light states emerge.

The former large $N$ limit, already considered in~\cite{Basile:2020mpt}, is considerably simpler, since it lies fully within the low-energy description, where the masses of KK states can be reliably computed. Dimensionally reducing the ten-dimensional gravitational EFT yields an effective action for the dilaton $\phi$ and the (canonically normalized) radion $\rho$~\cite{Basile:2020mpt,Basile:2021vxh}. The kinetic metric is canonical, and the masses of KK excitations around the $\ads_{p+2} \times \mathcal{M}_q$ flux compactifications scale according to
\begin{eqaed}\label{eq:kk_masses}
    m_\text{KK}^2 \propto M_{p+2}^2 \, e^{\frac{4}{\sqrt{p q}} \rho}
\end{eqaed}
in units of the \emph{dimensionally reduced} Planck mass~\cite{Antonelli:2019nar,Basile:2020mpt}. The vacuum values of the dilaton and radion pertaining to a given flux number $N$ are given by~\cite{Mourad:2016xbk,Antonelli:2019nar,Basile:2020mpt}
\begin{eqaed}\label{eq:dilaton_radion_vevs}
    \phi & \overset{N \gg 1}{\sim} - \, \frac{2}{(q-1) \gamma - 1} \, \log N \, , \\
    \rho & \overset{N \gg 1}{\sim} - \, \sqrt{\frac{q}{p}} \, \frac{4\gamma}{(q-1) \gamma - 1} \, \log N
\end{eqaed}
for large fluxes, so that the masses scale as the inverse radius of the internal space $\mathcal{M}_q$ measured in the $(p+2)$-dimensional Einstein frame~\cite{Antonelli:2019nar}. In particular, the squared KK masses scale as $N^{-3}$ and $N^{-2}$ for the orientifold models and the heterotic models respectively. 

Clearly, the KK tower becomes massless as $N \to \infty$. However, our main aim is to understand whether this limit lies at infinite distance, and whether the decay of KK masses is precisely exponential in the distance. Although the landscape of (metastable) vacua at stake is discrete, the flux tunneling process is realized by solitonic bubble profiles $\phi(r) \, , \, \rho(r)$ that continuously interpolate between the vacua. One could thus define a discrete-landscape distance via the metric in scalar field space defined by the effective action along such profiles.

This procedure would \emph{a priori} raise the issues of finding the relevant instanton solutions and minimizing the total distance along the many possible interpolations between two flux numbers $N_1 \, , \, N_2$. Actually, one can circumvent these obstacles bounding the distance from above and from below as follows, in such a way that the result is independent on the particular solitonic profile. From \cref{eq:dilaton_radion_vevs}, along any interpolating profile one can bound the metric via the inequalities
\begin{eqaed}\label{eq:dw_inequalities}
    \abs{\partial \phi} \, , \, \abs{\partial \rho} \leq \sqrt{(\partial \phi)^2 + (\partial \rho)^2} \leq \abs{\partial \phi} + \abs{\partial \rho} \, ,
\end{eqaed}
which imply that the distance between $N_1$ and $N_2$ is bounded by
\begin{eqaed}\label{eq:dw_distance_bound}
    a \, \log \abs{\frac{N_1}{N_2}} \leq \int \sqrt{(\partial \phi)^2 + (\partial \rho)^2} \, dr \leq b \, \log \abs{\frac{N_1}{N_2}} \, ,
\end{eqaed}
where $a \, , \, b$ are the $\mathcal{O}(1)$ constants that arise from \cref{eq:dilaton_radion_vevs}. Therefore the distance scales logarithmically, and accordingly the tower of KK masses decays exponentially fast in the distance, with a rate bounded by \cref{eq:dw_distance_bound} combined with \cref{eq:kk_masses}. In addition to supporting the distance conjecture in the absence of supersymmetry, as we have discussed this result resonates with the absence of scale separation in $\ads$ vacua~\cite{Gautason:2015tig, Lust:2019zwm, Lust:2020npd, Marchesano:2020qvg, DeLuca:2021mcj, Basile:2021vxh, Cribiori:2021djm}. Furthermore, since the KK masses are proportional to a positive power of the $\ads$ cosmological constant, the $\ads$ version of the distance conjecture~\cite{Lust:2019zwm} holds as well.

\begin{figure}[!ht]
\centering
\includegraphics[width=.4\textwidth]{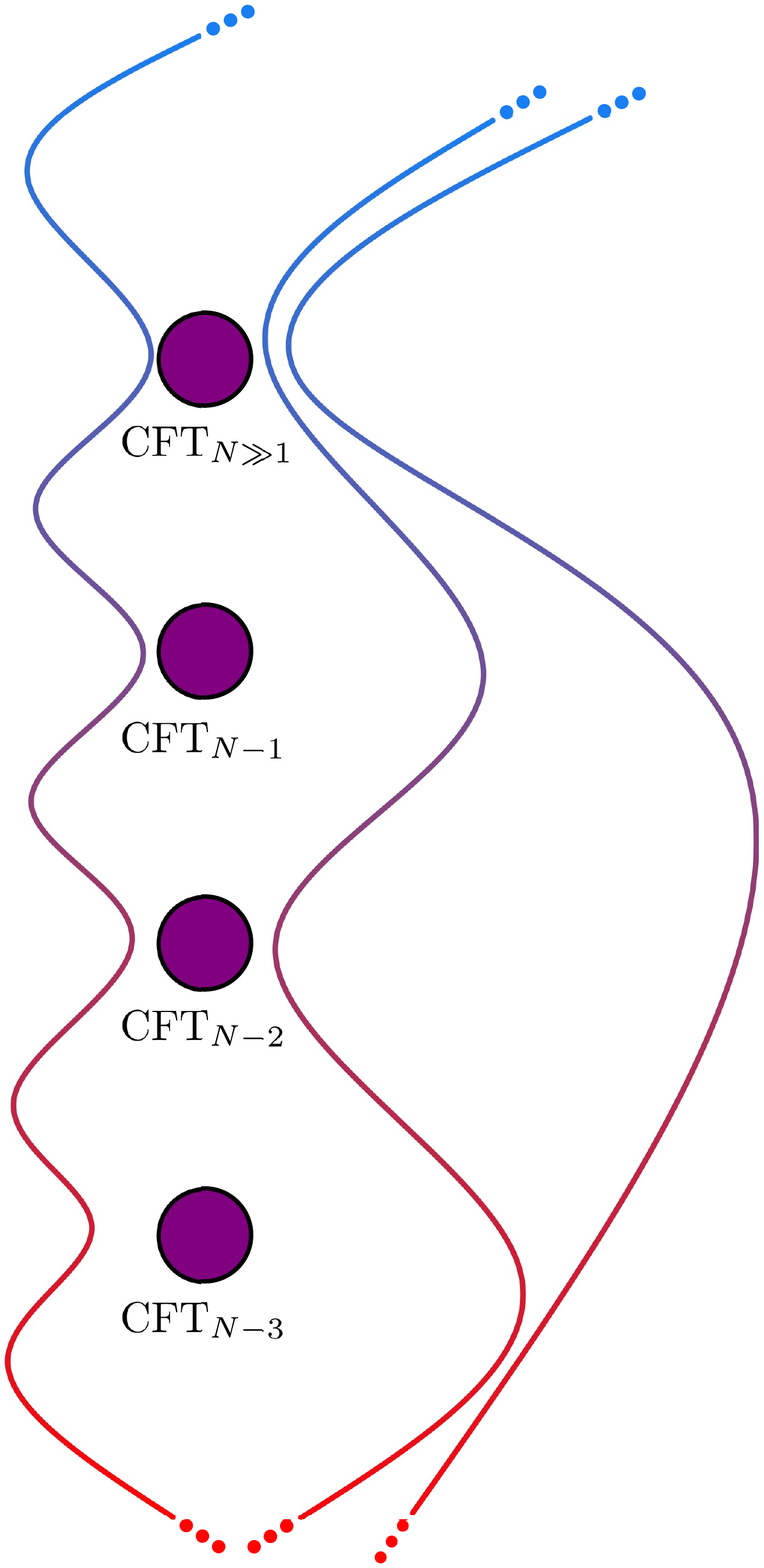}
\caption{The proposed holographic dual of the cascade of flux tunneling processes in the gravitational EFT is an RG flow in the boundary field theory~\cite{Antonelli:2018qwz}. Depending on the size, location and number of nucleation events, the trajectory can vary, approaching different fixed points. As $N \gg 1$ increases, the flows ought to approach the fixed points more closely, since the dual $\ads$ vacua are closer to stability~\cite{Antonelli:2019nar}.}
\label{fig:bubble-RG}
\end{figure}

Let us now focus on the opposite limit where $N = \mathcal{O}(1)$ is small. The EFT description is not trustworthy in this regime, since the string coupling and the curvatures are not negligible. In order to obtain an alternative description we appeal to holography, and in particular to the proposal of~\cite{Antonelli:2018qwz}~\footnote{See also~\cite{Kiritsis:2016kog, Ghosh:2017big, Gursoy:2018umf, Ghosh:2018qtg, Ghosh:2021lua} for discussions of a related proposal.}: the cascade of tunneling processes would be dual to a RG flow in the dual field theory, as depicted in \cref{fig:bubble-RG}. Importantly, the flows can only approach a fixed point for large $N$, since the $\ads$ vacua become more stable in this limit. As we shall find, the flow eventually reaches an endpoint for $N=1$, where for the BSB Sugimoto model supersymmetry is actually recovered.

\begin{figure}[!ht]
\centering
\includegraphics[width=.4\textwidth]{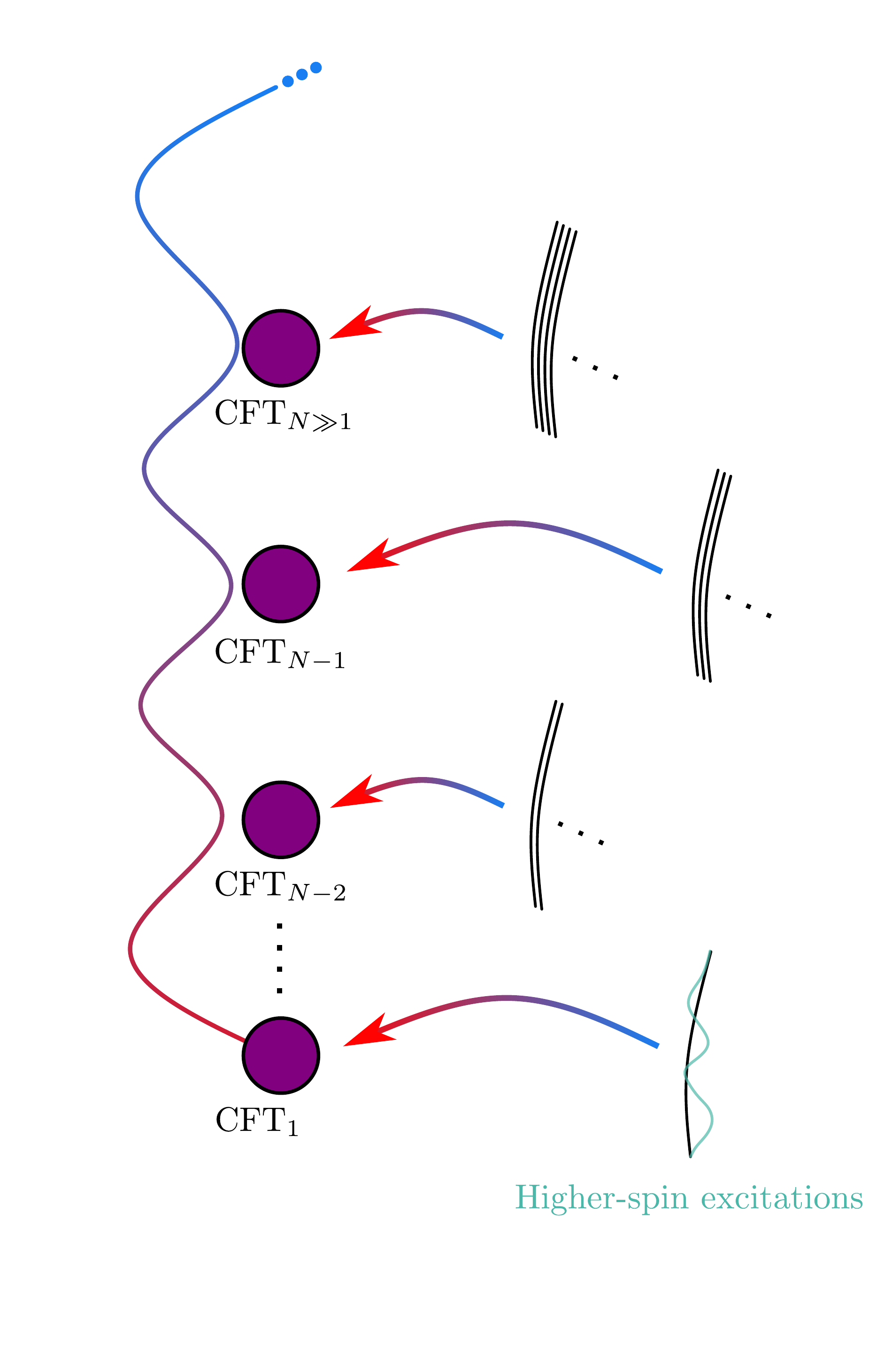}
\caption{The fixed points approached by the holographic RG flow can arise from the IR dynamics of the worldvolume gauge theory living on D1-brane stacks. The final state corresponds to the IR dynamics of a single D1-brane, which features a free sector with conserved single-trace higher-spin currents dual to massless single-particle higher-spin states. Furthermore, the Sugimoto model of~\cite{Sugimoto:1999tx} features emergent supersymmetry on account of $\text{Spin}(8)$ triality.}
\label{fig:bubble-RG_strings-fp}
\end{figure}

In the string models that we consider, the dual conformal field theory (CFT) for each $N$ ought to arise from the infrared (IR) regime of the worldvolume gauge theory that lives on a stack of $N$ parallel branes, following the original construction of~\cite{Maldacena:1997re}, and suitable deformations encode flux tunneling. In particular, in the remainder of this paper we shall focus on D1-branes in the orientifold models, since heterotic NS5-branes are considerably more difficult to deal with in this respect. Moreover, since the resulting field theories are two-dimensional, it is conceivable that progress can be achieved despite the absence of supersymmetry. As a minor technicality, we shall consider D1-branes in the parametrically controlled and flat region of the Dudas-Mourad background~\cite{Basile:2021mkd, Basile:2021vxh}, since there is no ten-dimensional Minkowski background. This holographic setup is shown in \cref{fig:bubble-RG_strings-fp}, where now each CFT arises from a worldvolume gauge theory in the IR, and the repulsion between branes triggers an RG flow. The corresponding operators arise by integrating out the brane separation modes, analogously to the more familiar case of supersymmetric D3-branes, where Higgsing generates Born-Infeld operators~\cite{Gubser:1998kv, Liu:1999kg, Intriligator:1999ai, Costa:1999sk, Costa:2000gk, Rastelli:2000xj, Caetano:2020ofu}.

The massless field content of the worldvolume gauge theory of $N$ D1-branes in the Sugimoto model comprises~\cite{Sugimoto:1999tx, Dudas:2001wd}~\footnote{See also~\cite{Gava:1998sv} for an analogous discussion of type I D1-branes.} a $\text{USp}(2N)$ gauge field $A$, scalars $X^i$ in the vector representation $\mathbf{8}_v$ of the transverse isometry group $\text{SO}(8)$ and the \emph{antisymmetric} representation of the gauge group and Majorana-Weyl fermions $\psi_+ \, , \psi_-$ in the spinor representations $\mathbf{8}_s , \mathbf{8}_c$ of $\mathfrak{so}(8)$ belonging to the symmetric ($\psi_+$) and antisymmetric ($\psi_-$) representations of the gauge group. Finally, bi-fundamental $\text{USp}(32) \times \text{USp}(2N)$ fermions $\lambda_-$ arise from the $\text{D}1-\overline{\text{D}9}$ open-string sector. On the other hand, in the type $0'\text{B}$ model a stack of $N$ D1-branes carries a $U(N)$ gauge group, and the scalars $X^i$ are in the \emph{adjoint}, while the Weyl fermion representations are unchanged~\cite{Dudas:2000sn}. One can verify that the worldvolume gauge anomalies cancel, since for the characteristic classes pertaining to the (anti)symmetric representations~\cite{Dudas:2000sn}
\begin{eqaed}\label{eq:anti-symmetric_classes}
    \text{Tr}_{S \, , \, A}F^2 = \left(N \pm 2 \right) \text{Tr}_\text{fund} F^2 + \left(\text{Tr}_\text{fund} F \right)^2 \, ,
\end{eqaed}
and thus one obtains a net contribution of $\frac{8 \times 4}{2} - 16 = 0$ from the chiral fermions. However, the gravitational anomaly does not cancel on the worldvolume without an inflow mechanism, which suggests that the theories are, in general, gapless~\cite{Delmastro:2021otj}. This is indeed what we find for the theories at stake. Two-dimensional gauge theories are amenable to a variety of methods, including light-cone techniques~\cite{Pauli:1985ps,Brodsky:1997de}, such as those pioneered in the original 't Hooft model~\cite{tHooft:1974pnl}, and (chiral) bosonization~\cite{Witten:1983ar, Gepner:1984au, Chung:1992mj,Frishman:1997uu}. The presence of scalars complicates matters to some extent, but one can expect that, due to the quartic potential, the IR dynamics be described by a non-linear $\sigma$ model (NL$\sigma$M)~\cite{Gava:1998sv,Bonati:2021oqq}. We shall return to this point later. For the time being, we shall focus on the last step of the cascade of decays, $N=2 \to N=1$. Since we are interested in the asymptotic behavior of the distance, this is the only relevant step of the RG flow. We shall begin from the endpoint $N=1$ itself, the CFT dual to the final state of the tunneling process. 

\subsection{Emergent supersymmetry and WZW cosets}\label{sec:wzw_cosets}

Among the various simplifications, crucially for $N=1$ the scalars \emph{decouple}, as we shall now explain. This is expected, since they describe transverse fluctuations of a single D1-brane in spacetime. As we shall see shortly, the remaining degrees of freedom in the worldvolume theories ought to flow to WZW coset models~\cite{Kutasov:1994xq,Isachenkov:2014zua,Dempsey:2021xpf,Delmastro:2021otj}, which can be constructed by means of non-Abelian bosonization. Moreover, they contain adjoint matter, whose IR coset structure is quite rich~\cite{Dalley:1992yy,Bhanot:1993xp,Boorstein:1993nd,Kutasov:1993gq,Gross:1997mx,Katz:2013qua,Dubovsky:2018dlk,Cherman:2019hbq,Komargodski:2020mxz}.

In detail, in the Sugimoto model the scalars and the fermions in the antisymmetric representation decouple because they belong to a singlet. Furthermore, the scalar potential and the Yukawa term vanish identically. Therefore, a free sector comprised of 8 pairs of a real scalar and a chiral fermion appears. These fields rearrange into a $\mathcal{N} = (0,1)$ Wess-Zumino multiplet, displaying \emph{emergent supersymmetry}~\cite{Goddard:1984hg,Friedan:1984rv,Lee:2010fy} at the endpoint of the tunneling cascade. This resonates with the considerations of~\cite{Ooguri:2016pdq,Antonelli:2018qwz}, and shows that the $N=1$ configuration is the stable final state of this process, protected by the restored supersymmetry. Let us observe that this remarkable phenomenon occurs due to $\text{Spin}(8)$ \emph{triality} (see~\cite{Tong:2019bbk} for a recent review), by virtue of the isomorphism $\mathbf{8}_v \simeq \mathbf{8}_s \simeq \mathbf{8}_c$. It would be interesting to explore its connections, if any, with the triality investigated in~\cite{Franco:2021ixh, Franco:2021vxq}. This remarkable occurrence in turn requires eight transverse dimensions. In the present setting this is only possible for (D-)\emph{strings}, realizing the proposal of~\cite{Lee:2018urn, Lee:2019xtm, Lee:2019wij} in a novel and peculiar fashion. Indeed, as we shall see, emergent strings are also the only case in which the (generalized) distance to the $N=1$ configuration is infinite. On the other hand, the antisymmetric fermions in the type $0'\text{B}$ model disappear for $N=1$, and thus only a non-supersymmetric free-boson CFT decouples. This is compatible with the non-supersymmetric origin of the theory. We now turn to the sectors that remain coupled to the gauge field.

\begin{figure}[!ht]
\centering
\includegraphics[width=.4\textwidth]{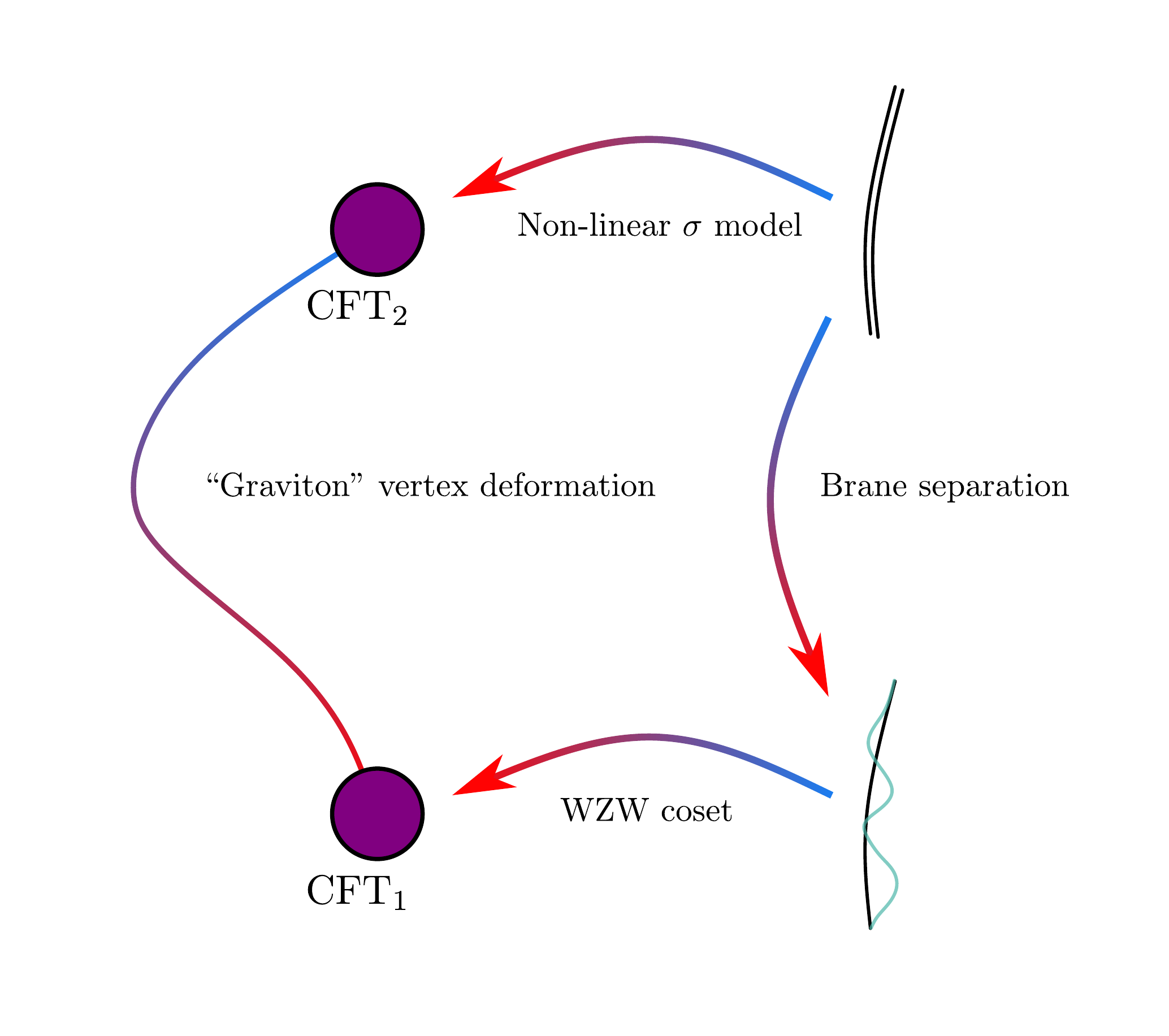}
\caption{The IR dynamics of the worldvolume gauge theories living on $N=2$ and $N=1$ D1-branes can be described via NL$\sigma$M and WZW coset constructions. The RG flow connecting the corresponding CFTs is triggered by the target-space metric, which is marginally irrelevant in the IR and yields an infinite distance along the flow.}
\label{fig:bubble-RG_square}
\end{figure}

For the Sugimoto model, the remaining degrees of freedom rearrange into the chiral WZW coset
\begin{eqaed}\label{eq:sugimoto_coset}
    \frac{\text{SO}(8\times 3)_1}{\text{SU}(2)_{8\times 2}} \times \frac{\text{SO}(16\times 2)_1}{\text{SU}(2)_{16 \times 1}} \, ,
\end{eqaed}
whose central charges read
\begin{eqaed}\label{eq:wzw_central_charges}
    \left(c_{\text{L}}^{\text{IR}} \, , \, c_{\text{R}}^{\text{IR}}\right) & = \left(c_{\text{L}}^{\text{UV}} \, , \, c_{\text{R}}^{\text{UV}}\right) - \left(\frac{8}{3} \, , \, \frac{8}{3} \right) \\
    & = \left(12 - \, \frac{8}{3} \, , \, 16 - \, \frac{8}{3} \right)
\end{eqaed}
compatibly with the same gravitational anomaly $c_\text{R} - c_\text{L} = 4$. Interestingly, the \emph{total} central charges of this model, including the free $\mathcal{N} = (0,1)$ SCFT sector, can be recast in the form
\begin{eqaed}\label{eq:wzw_central_charges_s-dual}
    \left(12 \, , \, 24\right) + \left(\frac{16}{3} \, , \, \frac{4}{3} \right) \, ,
\end{eqaed}
which is tempting to identify with the central charges of (the transverse degrees of freedom of) a dual ten-dimensional heterotic string, plus a ``correction'' due to supersymmetry breaking. While this is at best an amusing hint, the prospect of a strong-weak duality in the absence of supersymmetry remains tantalizing~\cite{Blum:1997cs,Blum:1997gw,Faraggi:2007tj,Faraggi:2009xy}, and we shall elaborate on this point shortly. Furthermore, the correction to the left-moving supersymmetric sector is compatible with a Gepner model built by two copies of the $k = 16$, $\mathcal{N} = 2$ minimal models. The deviation
\begin{eqaed}\label{eq:wzw_central_charges_free}
    \left(8 \, , \, 12\right) + \left(\frac{28}{3} \, , \, \frac{40}{3} \right)
\end{eqaed}
from the free $\mathcal{N} = (0,1)$ SCFT is also compatible with Gepner models of this type. The relation between this WZW coset model and the $N=2 \to N=1$ transition is depicted in \cref{fig:bubble-RG_square}.

These considerations on the worldsheet theory of a D-string in the Sugimoto model, which answer the question originally posed in~\cite{Sugimoto:1999tx}, can be complemented by arguments based on the low-energy effective action, in the spirit of the supersymmetric heterotic-type I duality~\cite{Polchinski:1995df}. Performing a na\"{i}ve S-duality transformation on the spacetime metric and dilaton fields, the tadpole potential of \cref{eq:tadpole_potential} translates into the string-frame contribution $e^{-4\phi} = e^{-2\phi} \times e^{-2\phi}$. Intriguingly, this structure mirrors the presence of two decoupled sectors of the worldsheet theory: quantizing it on disjoint unions of two Riemann surfaces, with the free geometric sector on one connected component and the non-geometric WZW coset on the other, would seem to reproduce this effect while being consistent with a single heterotic-like string in physical spacetime. In particular, the leading contribution to string perturbation theory would stem from a surface with topology $S^2 \sqcup S^2$, and scale precisely as $g_s^{-2} \times g_s^{-2}$. Notably, only one sector is geometric, preserving the standard interpretation of a connected worldsheet in physical spacetime, while the other sector is non-geometric. Let us emphasize that disconnected worldsheets have already appeared in the literature, in the context of D-instantons and non-perturbative effects~\cite{Polchinski:1994fq, Green:1995mu, deBoer:1998gyt}. All in all, the emergent superconformal free heterotic-like sector, together with a non-geometric sector and the corresponding leading contribution to an S-dual EFT seem to point to a novel heterotic-orientifold duality, although the arguments that we have presented are but compelling indications for the time being. As we shall discuss in the following section, this configuration arises from an infinite-distance emergent-string limit, which is also a smoking gun of dualities of this type~\cite{Ooguri:2006in}. It would be interesting to attempt to construct a Polyakov quantization of such a heterotic string.

For the type $0'\text{B}$ model, the $N=1$ worldvolume gauge theory is Abelian, and bosonization simplifies accordingly~\cite{Kutasov:1993gq, Gepner:1984au} in a slight generalization of the (chiral) Schwinger model~\cite{Schwinger:1962tn,Schwinger:1962tp,Jackiw:1984zi}. The upshot is that a single linear combination of the bosonized chiral scalars becomes gapped in the IR, resulting in one less massless fermion for both chiralities~\cite{Kutasov:1993gq}~\footnote{See also~\cite{Razamat:2020kyf, Tong:2021phe} for related discussions on mass generation in chiral gauge theories.}. Since the photon acquires a dynamical mass, the resulting CFT is free. It would be interesting to assess whether emergent supersymmetry can arise in this model in terms of two-dimensional $\mathcal{N} = (0,1)$ Fermi supermultiplets.

\section{Infinite distances and emergent strings}\label{sec:infinite_distances}

Having described the endpoint of the tunneling process holographically, we can now turn to the RG flow. In both orientifold models, the CFT dual to the final state contains a free sector, and thus single-trace higher-spin currents that are conserved. On general grounds, one expects that the higher-spin symmetry be broken along the flow, and that the anomalous dimensions vanish continuously at the endpoint. Similarly to our analysis of KK masses, one is thus led to seek a suitable generalization of the Zamolodchikov metric, since there is no exact conformal manifold. Such a notion was introduced by O'Connor and Stephens on information-theoretic grounds~\cite{oconnor}, explored by Dolan~\cite{Dolan:1993cf, Dolan:1995zq} and subsequently revisited in numerous settings, most recently by Stout in the context of the distance conjecture~\cite{Stout:2021ubb}~\footnote{See~\cite{Lassig:1989tc, Kar:2001qm, Kar:2002wx, Beny:2012qh, Jack:2013sha, Maity:2015rfa} for related discussions on the geometry of the RG flow, and~\cite{Provost:1980nc, Ruppeiner:1995, Ruppeiner:1996} for discussion of the role of the information metric in quantum systems.}.

Parametrizing the theory space by operators $\mathcal{O}_a(x)$ with couplings $\lambda_a$, the metric reads
\begin{eqaed}\label{eq:sos_metric}
    g_{ab}(\lambda) = \int d^dx \, \langle \mathcal{O}_a(x) \mathcal{O}_b(0) \rangle_\lambda
\end{eqaed}
up to a volume factor, in a special coordinate system in which the action is linear in the couplings, so that $g$ is also the negative Hessian of the vacuum energy density~\cite{oconnor}. While this metric was developed applying the tools of quantum information theory to RG flows, a similar metric was defined by Anselmi in the context of Lorentz-breaking field theories~\cite{Anselmi:2011bp}~\footnote{See also~\cite{Anselmi:1997rd, Anselmi:1999xk, Anselmi:2000fu} for earlier works in a similar spirit in holographic settings.}. This metric trades the integral over spacetime in \cref{eq:sos_metric} for an energy scale $\mu$, which appears in a timelike position argument $x_t \equiv \mu^{-1} (1, \mathbf{0})$. The metric reads      
\begin{eqaed}\label{eq:ans_metric}
    g_{ab}(\lambda) = \abs{x_t}^{2d} \langle \mathcal{O}_a(x_t) \mathcal{O}_b(0) \rangle_\lambda \, ,
\end{eqaed}
and in both metrics the correlators are connected. In the following we shall compute distances along the RG flow using both metrics. While the results coincide up to a constant in this case, the metrics differ in general.

As we have anticipated, it is natural to expect that the IR regime of the RG flow be dominated by a NL$\sigma$M deformation of the free (S)CFT as shown in \cref{fig:bubble-RG_square}. Indeed, in two dimensions these are the only (classically) marginal ones, since the IR fixed point is Gaussian and classical power counting applies. On the one hand, from the point of view of the renormalization group, these deformations are present during a generic flow. On the other hand, as we shall discuss, they arise from the repulsion of branes in the specific models that we consider. In fact, such deformations are marginally irrelevant, and the distance along the flow can diverge only in this case. This stems from the exponential decay $\sim e^{- \Delta \, t}$ of strictly irrelevant deformations as the RG time $t \to +\infty$ in the IR, so that the integral $\int^\infty dt$ computing the total distance converges. Once again, this peculiar fact dovetails nicely with the emergent string proposal of~\cite{Lee:2018urn, Lee:2019xtm, Lee:2019wij}, namely the infinite distance limit at stake only exists when the extended objects that becomes tensionless is a (D-)string. Indeed, in contrast one can verify that, were an analogous Gaussian fixed point present for NS5-branes, there would not be any marginally irrelevant deformations preserving the symmetries, and hence no corresponding infinite-distance limit.

Here we shall focus on the bosonic sector, defined by the action
\begin{eqaed}\label{eq:free_cft}
    S_\text{CFT} = \frac{1}{2\pi \alpha} \int d^2z \, \delta_{ij} \, \partial X^i \overline{\partial} X^j
\end{eqaed}
for a suitable constant $\alpha$~\footnote{Normally, $\alpha$ has dimensions of squared length. In order to keep \cref{eq:general_correlator} free of unnecessary clutter, we work in string units for the $X^i$, so that $\alpha$ is dimensionless.}, although by power counting one does not expect that the fermionic sector yield different results. The deformation of the NL$\sigma$M effectively turns on the tension of the D-string, along the lines of~\cite{Gaberdiel:2013jpa, Gaberdiel:2015uca}~\footnote{See also~\cite{David:1999ec, Avery:2010er, Avery:2010hs, Avery:2010vk} for analogous considerations in the context of black holes.}, and takes the general form
\begin{eqaed}\label{eq:deformation_metric}
    V & = \frac{1}{2\pi \alpha} \int d^2z \, : h_{ij}(X) \, \partial X^i \overline{\partial} X^j : \\
    & \equiv \frac{1}{2\pi \alpha} \int d^2z \, \mathcal{V}(z) \, ,
\end{eqaed}
where we shall take the
\begin{eqaed}\label{eq:h_fourier}
    h_{ij}(X) = \int \frac{d^Dk}{(2\pi)^D} \, f_{ij}(k) \, e^{ik \cdot X}
\end{eqaed}
with $D=8$ transverse target-space dimensions. In string perturbation theory, \cref{eq:deformation_metric} would be a graviton vertex operator. While this terminology is useful, one ought to keep in mind that in this holographic CFT its meaning is different, and in particular there is no on-shell constraint. Although \cref{eq:deformation_metric} is quite general, we shall derive a more concrete expression for the string models that we consider.

The corresponding one-loop Ricci flow~\cite{Friedan:1980jf,Hamilton:1982gg, Callan:1985ia, Callan:1989nz}~\footnote{See also~\cite{DeBiasio:2020xkv, DeBiasio:2021yoe} for recent considerations on geometric flows in the context of the swampland, albeit from a different angle.} for transverse-traceless $h_{ij} \ll 1$ is
\begin{eqaed}\label{eq:ricci_flow}
    \frac{d}{dt} \, h_{ij} = - \, \alpha \, R_{ij} \sim \frac{\alpha}{2} \, \Box h_{ij} \, ,
\end{eqaed}
so that the Fourier modes of $h_t$ run according to $f_{ij}(k) \, e^{- \frac{\alpha k^2}{2} t}$. Thus, the perturbation of \cref{eq:deformation_metric} is indeed marginally irrelevant in the IR. In the ensuing discussions we shall consider modes of this type, in order to simplify the computations. However, the qualitative results should be unaffected in the general case.

The ``Stout-O'Connor-Stephens'' (SOS) quantum information distance along the RG flow in the deep IR is then asymptotically given by
\begin{eqaed}\label{eq:sos_distance}
    d\ell^2 \overset{\text{IR}}{\sim} \frac{1}{(2\pi \alpha)^2} \int d^2z \, \langle \partial_t\mathcal{V}(z) \partial_t\mathcal{V}(0) \rangle_0 \, dt^2 \, ,
\end{eqaed}
where the subscript indicates free correlators, since the theory is free in the IR. In the following, all asymptotic signs refer to the IR limit $t \to +\infty$. To compute the correlator in \cref{eq:sos_distance}, we shall employ the general Gaussian formula
\begin{eqaed}\label{eq:general_correlator}
    \langle \prod_{n} e^{ip_n \cdot X(z_n,\overline{z}_n)} e^{i \int S \cdot X} \rangle_0 & = \delta\!\left(\sum_n p_n\right)\!\! \prod_{n < m} \!\abs{z_n-z_m}^{\alpha \, p_n \cdot p_m} \\ 
    & \times \, e^{\frac{\alpha}{2} \, \int S \cdot G \cdot S + \alpha \sum_n p_n (\int G \cdot S)(z_n, \overline{z}_n)}
\end{eqaed}
for free correlators, where normal ordering is understood and $G(z,w) \equiv \log\abs{z-w}$. Functionally differentiating with respect to the auxiliary source $S(z,\overline{z})$ and differentiating with respect to worldsheet position, one can obtain normal-ordered correlators subtracting the contact terms and setting the insertion points equal at the end of the calculation. Applying this technique to \cref{eq:sos_distance}, taking into account the connected contributions and the transverse-traceless ``graviton'', one finds
\begin{eqaed}\label{eq:unintegrated_correlator_computation}
    \langle \partial_t\mathcal{V}(z) \partial_t\mathcal{V}(0) \rangle_0 \sim \frac{\alpha^4}{16}\int \frac{d^Dk}{(2\pi)^D} \, \abs{z}^{-\alpha k^2-4} k^4 \, \norm{f_t(k)}^2
\end{eqaed}
in the IR, where $\norm{f_t(k)}^2 \equiv f_{ij}(k)\overline{f}^{ij}(k) \, e^{-\alpha k^2 t}$. Therefore, as $t \to +\infty$ the relevant contributions arise from the $k \to 0$ region of integration. The integrated correlator thus has the IR behavior
\begin{eqaed}\label{eq:integrated_correlator_IR}
    \int d^2z \, \langle \partial_t\mathcal{V}(z) \partial_t\mathcal{V}(0) \rangle_0 \sim \frac{\pi\alpha^4}{48a^2}\int \frac{d^Dk}{(2\pi)^D} \, k^4 \, \norm{f_t(k)}^2 \, ,
\end{eqaed}
where we have evaluated the integral at $k=0$ using a lattice regulator~\cite{Amit:1979ab, Das:1986da, Watabiki:1987qg, Sathiapalan:1987vj}~\footnote{The same results can be obtained with other regularizations.}, which here amounts to the replacement $\frac{1}{z} \to \frac{\overline{z}}{\abs{z}^2+a^2}$. The resulting asymptotic SOS distance is
\begin{eqaed}\label{eq:sos_distance_computation}
    d\ell^2 \overset{\text{IR}}{\sim} \frac{\alpha^2}{192 \pi a^2}\int \frac{d^Dk}{(2\pi)^D} \, k^4 \, \norm{f(k)}^2 \, e^{- \alpha k^2 t} \, dt^2 \, ,
\end{eqaed}
where the dependence on the RG time $t$ has been factored out for clarity. Taking into account the volume factor in \cref{eq:sos_metric}, this differs by a factor of $a^{-2}$ from the ``intensive'' metric of~\cite{Stout:2021ubb}, which is thus finite. At this point, the $k \to 0$ asymptotics of the Fourier modes of $h$ is needed in order to evaluate the IR distance. As we shall see, it seems reasonable to assume that, schematically, $f(k) \sim \abs{k}^{-m}$ as $k \to 0$, for some $m>0$. Indeed, we shall shortly verify this assumption in the string models that we consider. Then, up to an irrelevant constant,
\begin{eqaed}\label{eq:sos_distance_asymptotics}
    d\ell^2 & \sim \int_0^\infty dk \, k^{D+3-2m} \, e^{- \alpha k^2 t} \, dt^2 \sim \frac{dt^2}{t^{\frac{D}{2}+2-m}} \, ,
\end{eqaed}
so that the distance is infinite insofar as $m \geq \frac{D}{2} = 4$.

As one can see from \cref{eq:unintegrated_correlator_computation}, the metric of \cref{eq:ans_metric} actually coincides with the SOS metric up to a constant, since multiplying by $\abs{z}^4$ leaves $\abs{z_t}^{- \alpha k^2} = a^{- \alpha k^2} \, e^{- \alpha k^2 t}$ when evaluated at the RG-scale insertion $z_t$. This reconstructs the RG flow of the ``graviton vertex'' operator, as expected from the Callan-Symanzik equation.

\subsection{Graviton vertex from brane separation}\label{sec:brane_separation}

From the preceding discussion, one can expect that the IR regime of the $N=2$ configuration be encoded, at least partly, in the NL$\sigma$M described by the minima of the scalar potential, \emph{i.e.} mutually commuting matrices. In order to describe the geometry of the resulting manifold, let us begin from the simpler case of the type $0'\text{B}$ model, where the scalars belong to the adjoint representation of $U(2)$. Mutually commuting Hermitian matrices $X^i$ can be parametrized in terms of their eigenvalues $\Lambda^i = \text{diag}(x^i_1 \, \dots \, x^i_N)$ and of a unitary matrix $U$ that simultaneously diagonalizes them according to $X^i = U \, \Lambda^i \, U^\dagger$. The canonical kinetic-term metric $\text{Tr}(dX^i dX^i)$ is thus pulled back to
\begin{eqaed}\label{eq:un_worldvolume_metric}
    ds^2_{U(N)} = \sum_k d\mathbf{x}_k^2 + 2 \sum_{p < q} \norm{\mathbf{x}_p - \mathbf{x}_q}^2 \, \text{Tr} \, \theta \theta^\dagger \, ,
\end{eqaed}
where $\mathbf{x}_k = (x^i_k)$ can be interpreted as the transverse position vector of the $k$th brane and $\theta = U^\dagger dU$ is the Maurer-Cartan form. For $N=2$ one can express \cref{eq:un_worldvolume_metric} in terms of the center-of-mass and relative positions $\mathbf{x}_c = \frac{\mathbf{x}_1 + \mathbf{x}_2}{\sqrt{2}}$, $\mathbf{r} = \frac{\mathbf{x}_1 - \mathbf{x}_2}{\sqrt{2}}$, which yields
\begin{eqaed}\label{eq:u2_worldvolume_metric_rel}
    ds^2_{U(2)} = d\mathbf{x}_c^2 + d\mathbf{r}^2 + 4 \, \mathbf{r}^2 \, \text{Tr} \, \theta \theta^\dagger \, .
\end{eqaed}
Integrating out the center-of-mass position does not affect the relative dynamics, while integrating out $U$, which encodes the interaction between the branes in this sector, generates an effective action for $\mathbf{r}$ that describes fluctuations of the remaining brane. We are interested in the target-space metric, encoded in the kinetic term, at large brane separation $\norm{\mathbf{r}}$, which translates into the $k \to 0$ asymptotics of \cref{eq:sos_distance_asymptotics}. Parametrizing $U$ with local coordinates $u^a$ and writing the kinetic term corresponding to the last term of \cref{eq:u2_worldvolume_metric_rel} as
\begin{eqaed}\label{eq:unitary_sigma-model}
    e^{2 \log \norm{\mathbf{r}}} \, G_{ab}(u) \, \partial u^a \cdot \partial u^b \, ,
\end{eqaed}
up to a constant, one recognizes a NL$\sigma$M coupled to a ``dilaton'' $\Phi = - \log \norm{\mathbf{r}}$ in the sense of~\cite{Gusev:1999cv}. Large separations intuitively correspond to the semiclassical limit, since the branes interact weakly, and thus a one-loop analysis is expected to be reliable. In order to see this more clearly, one can choose normal coordinates to perform a covariant background-field expansion~\cite{Ecker:1971xko,Friedan:1980jf,Flore:2012ma} about a point,
\begin{eqaed}\label{eq:covariant_expansion}
    G_{ab}(u) = \delta_{ab} - \, \frac{1}{3} \, R_{acbd}(0) \, u^c \, u^d + \, \dots
\end{eqaed}
which we take as the origin of coordinates. One can then canonically redefine $u^a = e^{- \log \norm{\mathbf{r}}} \, \tilde{u}^a$ to absorb the ``dilaton'' in the quadratic term in the fluctuations, while all the other terms are suppressed in the large $\norm{r}$ limit. As a result, one can indeed perform a one-loop computation along the lines of~\cite{Gusev:1999cv}. Integrating by parts one can recast the quadratic term as
\begin{eqaed}\label{eq:quadratic_term}
    \tilde{u}^a \left(- \, \Box - \, \Box \Phi + (\partial \Phi)^2\right) \delta_{ab} \, \tilde{u}^b \, ,
\end{eqaed}
so that the heat-kernel expansion~\cite{Gusev:1999cv, Vassilevich:2003xt, Flore:2012ma} yields a single local two-derivative term that corrects the field-space metric. This term is proportional to the first heat-kernel coefficient $a_2$, and the corrected metric reads
\begin{eqaed}\label{eq:corrected_metric}
   \left( \delta_{ij} - \, \frac{r_i \, r_j}{\pi \, r^4} \, \log \frac{L}{a} \right) dr^i dr^j
\end{eqaed}
with IR and UV cutoffs $L \, , \, a$.

For the Sugimoto model the scalars belong to the antisymmetric representation of $\text{USp}(4) \simeq \text{Spin}(5)$, and mutually commuting matrices of this type can be parametrized by a rotation $R \in \text{SO}(4)$ and block-diagonal matrices built as linear combinations $\Lambda^j = i \, \frac{x^j_k}{2} \, \Omega_k$, where $k = 1 \, , \, 2$ and
\begin{eqaed}\label{eq:omega_matrices}
    \Omega_1 & = \mqty(1 & 0 \\ 0 & 0) \otimes \mqty(0 & 1 \\ -1 & 0) \, , \\
    \Omega_2 & = \mqty(0 & 0 \\ 0 & 1) \otimes \mqty(0 & 1 \\ -1 & 0) \, .
\end{eqaed}
Then, writing $X^i = R \, \Lambda^i \, R^{-1}$ the canonical metric $\text{Tr}(dX^i dX^i)$ pulls back to
\begin{eqaed}\label{eq:usp4_metric}
    ds^2_{\text{USp}(4)} = \sum_k d\mathbf{x}^2_k - \, \frac{1}{4} \sum_{p, q} \, \mathbf{x}_p \cdot \mathbf{x}_q \, \text{Tr}[\Omega_p , \theta][\Omega_q , \theta] \, ,
\end{eqaed}
and one can check that the dependence on the center-of-mass and relative coordinates separate, with no mixed terms. The resulting expression for the trace in \cref{eq:usp4_metric},
\begin{eqaed}\label{eq:usp_4_metric_theta}
    & \mathbf{x}_0^2 \left( (\theta_{14} + \theta_{23})^2 + (\theta_{13} - \theta_{24})^2 \right) \\
    & + \mathbf{r}^2 \left( (\theta_{14} - \theta_{23})^2 + (\theta_{13} + \theta_{24})^2 \right) \, ,
\end{eqaed}
encodes two pull-backs on the hyperplanes
\begin{eqaed}\label{eq:hyperplanes}
    \theta_{12} = \theta_{34} = 0 \, , \quad \theta_{14} = \pm \, \theta_{23} \, , \quad \theta_{13} = \mp \, \theta_{24}
\end{eqaed}
in the space of antisymmetric matrices, since these combinations do not appear in \cref{eq:hyperplanes}. Hence, introducing local coordinates $u^a$ one arrives at kinetic terms of the type
\begin{eqaed}\label{eq:usp4_metric_sigma-model}
   \left(\mathbf{x}_0^2 \, G^{(0)}_{ab}(u) + \mathbf{r}^2 \, G^{(r)}_{ab}(u) \right) \partial u^a \cdot \partial u^b \, .
\end{eqaed}
Once again one can reabsorb the ``dilaton'' with a field redefinition $u^a = e^{- \log \norm{\mathbf{r}}} \, \tilde{u}^a$, and integrating out $\mathbf{x}_0$ yields terms that are subleading at large $\norm{\mathbf{r}}$. Repeating the above argument for the type $0'\text{B}$ model yields a corrected metric of the form of \cref{eq:corrected_metric}, albeit with a halved prefactor $\frac{\dim \text{SO}(4)-4}{4\pi} = \frac{1}{2\pi}$ in front of the logarithm due to the constraints of \cref{eq:hyperplanes}.

The scaling $r^{-2}$ of \cref{eq:corrected_metric} can be compared with a bulk calculation. In Poincar\'{e} coordinates, the near-horizon $\ads$ throat warp factor for D1-branes scales as $\frac{L^2}{z^2}$, and according to our setup one expects that the background metric be fixed by the D8-branes in the controlled region. As a result, when all of the branes have repelled each other, one expects the correction to the transverse-space metric of the remaining brane located at $\mathbf{x}$ to be well-approximated by a linear superposition of the form
\begin{eqaed}\label{eq:bulk_brane_metric}
    h_{ij}(r) & \sim \int_{S^7} \, \frac{d\Omega_7(\mathbf{n})}{\norm{\mathbf{x} - r \, \mathbf{n}}^2} \, \delta_{ij} \sim \frac{\delta_{ij}}{r^2}
\end{eqaed}
for large separation $r$, which reproduces the overall scaling $r^{-2}$ of the correction in \cref{eq:corrected_metric}.

All in all, comparing with \cref{eq:sos_distance_asymptotics} one finds
\begin{eqaed}\label{eq:stringy_graviton_m}
    m = D - 2 = 6 \, ,
\end{eqaed}
from which the asymptotic distance is proportional to the RG time,
\begin{eqaed}\label{eq:stringy_distance_asymptotics}
    d\ell^2 \sim \lambda^2 \, \frac{dt^2}{t^{\frac{8}{2} + 2 - 6}} = \lambda^2 dt^2 \, ,
\end{eqaed}
where the (regularization-dependent, but calculable) proportionality constant $\lambda$ has been reinstated. As a result, exponential decay in $t$ in the IR is tantamount to exponential decay in the distance $\ell$ at large distances. This asymptotic scaling is crucial in order to establish such a behavior of anomalous dimensions as the distance diverges. We would like to stress that the result of \cref{eq:stringy_distance_asymptotics} is not generic in the space of marginally irrelevant deformations, rather it appears to be specific to our settings arising from string theory.

\subsection{Anomalous dimensions of scalar operators}\label{sec:anomalous_dimensions_scalar}

We are now ready to compute anomalous dimensions along the RG flow, using the scheme in \cref{eq:integrated_correlator_IR}. One can then translate the dependence on the RG time $t$, which appears consistently with the Callan-Symanzik equation, into a dependence on the distance $\ell$. The resulting behavior turns out to be precisely consistent with the distance conjecture, as we shall see below. Since out approach is holographic, we mostly refer to the CFT counterpart of the distance conjecture~\cite{Baume:2020dqd, Perlmutter:2020buo}, according to which, in two dimensions, the quantity of interest is the gap in the spectrum of scalar primaries $\mathcal{O}_p \equiv \; \frac{2}{\alpha} :e^{ip \cdot X}:$. In order to find a non-trivial scalar gap, one can compactify the target space in a $D$-torus with quantized ``momenta'' $p \, , \, q$. Then, we shall turn to single-trace higher-spin currents, which ought to describe higher-spin single-particle states in the bulk.

For scalar operators, first-order conformal perturbation theory suffices to obtain the leading-order asymptotics. The first-order contribution to the correlator of the scalar primaries is
\begin{eqaed}\label{eq:scalar_pert_correlator}
    \langle \mathcal{O}_p(z) \mathcal{O}_q(0) \rangle_{1} = \int \frac{d^2x \, p^i \, p^j \, \overline{f}_{ij}(p+q) \, \abs{z}^{\alpha p \cdot q+2}}{\abs{z-x}^{\alpha (p^2 + p \cdot q)+2} \abs{x}^{\alpha (q^2 + p \cdot q)+2}} \, ,
\end{eqaed}
and extracting the divergent part, which is logarithmic for small $p \, , \, q$ which have the least decreasing anomalous dimensions, entails separating the contributions of the integration regions $x \to 0$ and $x \to z$, according to
\begin{eqaed}\label{eq:scalar_div}
    \langle \mathcal{O}_p(z) \mathcal{O}_q(0) \rangle^{\text{div}}_{1} & = \int_z \frac{d^2x \, p^i \, p^j \, \overline{f}_{ij}(p+q) \, \abs{z}^{- \alpha q^2}}{\abs{z-x}^{\alpha (p^2 + p \cdot q)+2}} \\
    & + \int_0 \frac{d^2x \, p^i \, p^j \, \overline{f}_{ij}(p+q) \, \abs{z}^{- \alpha p^2}}{\abs{x}^{\alpha (q^2 + p \cdot q)+2}} \, .
\end{eqaed}
Evaluating the integrals with a lattice regulator, for small $p,q$ one finds the logarithmic divergence
\begin{eqaed}\label{eq:scalar_div_reg}
    &\abs{z}^{- \alpha (p+q)^2} \left(\left(\frac{\abs{z}}{a}\right)^{\alpha (p^2 + p \cdot q)} + \left(\frac{\abs{z}}{a}\right)^{\alpha (q^2 + p \cdot q)}\right) \\
    & \sim 2 \, \abs{z}^{- \alpha (p+q)^2} \left(1 + \frac{\alpha (p+q)^2}{2} \, \log \frac{\abs{z}}{a} \right) \sim \abs{z}^{- \alpha \frac{(p+q)^2}{2}}
\end{eqaed}
up to a scheme-dependent positive multiplicative constant and a factor of $p^i \, p^j \, \overline{f}_{ij}(p+q) \abs{z}^{\alpha p \cdot q}$. The prefactor reflects the tree-level correlator 
\begin{eqaed}\label{eq:scalar_tree_correlator}
    \langle \mathcal{O}_p(z) \mathcal{O}_q(0) \rangle_{0} = \frac{4}{\alpha^2} \, \delta(p+q) \, \abs{z}^{\alpha p \cdot q} \, ,
\end{eqaed}
while the exponent in \cref{eq:scalar_div_reg} reconstructs the one-loop Ricci flow evaluated at the RG time $t = \log \frac{\abs{z}}{a}$, as expected from the Callan-Symanzik equation
\begin{eqaed}\label{eq:callan-symanzik}
    \langle \mathcal{O}_p(s z) \mathcal{O}_q(0) \rangle_h = Z^{-1}_{pp'} \, Z^{-1}_{qq'} \, s^{\alpha p' \cdot q'}  \, \langle \mathcal{O}_{p'}(z) \mathcal{O}_{q'}(0) \rangle_{h(s)} \, .
\end{eqaed}
Therefore the (``matrix'' of) anomalous dimensions $\gamma_{p q}$, obtained differentiating the anomalous contribution with respect to $-t$, scales according to
\begin{eqaed}\label{eq:scalar_anomalous_dims}
    \gamma_{p q} & \sim \alpha \, \frac{(p+q)^2}{2} \, p^i p^j \, \overline{f}_{ij}(p+q) \, e^{- \alpha \frac{(p+q)^2}{2} t} \\
    & \sim p^i p^j \, \overline{R}_{ij}(p+q) \, e^{- \alpha \frac{(p+q)^2}{2} \lambda \ell}
\end{eqaed}
up to a (scheme-dependent) constant, where $R_{ij}(k) \sim \frac{k^2}{2} \, f_{ij}(k)$ denotes the Fourier modes of the (linearized) Ricci tensor. This expression shows that ``graviton'' zero-modes do not result in anomalous dimensions, which is indeed the case since they are exactly marginal deformations leaving the theory free.

All in all, \cref{eq:scalar_anomalous_dims} highlights an exponential decay of the scalar gap with the distance, on account of \cref{eq:stringy_distance_asymptotics}. As anticipated, this result supports the distance conjecture, and its various refinements, in the absence of (linear) supersymmetry. In particular, the emergent string scenario is realized in a novel fashion, since the $N=1$ configuration lies at infinite distance only in two dimensions.

\subsection{Anomalous dimensions of higher-spin currents}\label{sec:anomalous_dimensions_currents}

The free-boson CFT defined by the action of \cref{eq:free_cft} also possesses single-trace (anti-)holomorphic higher-spin currents of the form~\cite{Bakas:1990ry, Bergshoeff:1990cz, Bergshoeff:1991dz, Pope:1991ig, Hull:1991ca, Hull:1992hy, Anselmi:1998bh, Anselmi:1998ms, Skvortsov:2015pea, Giombi:2016hkj}
\begin{eqaed}\label{eq:free_hs_currents}
    J_s^{ij} = \sum_{n=1}^{s-1} (-1)^n \, A_n^s \, :\partial^{n}X^i \partial^{s-n}X^j:
\end{eqaed}
with suitable coefficients $A_k^s$~\cite{Bakas:1990ry, Bergshoeff:1990cz, Bergshoeff:1991dz, Pope:1991ig, Skvortsov:2015pea, Giombi:2016hkj, Gerasimenko:2021sxj}~\footnote{The coefficients can be also obtained expanding a bilocal expression in the fields~\cite{Gerasimenko:2021sxj}.} such that they generate a $W_{\infty}$ algebra. We shall focus on the $O(D)$ singlets $J_s \equiv \delta_{ij} \, J_s^{ij}$ for simplicity, which only exist for even $s$~\cite{Giombi:2016hkj}, retracing the computation for scalar operators. The leading-order correction to the correlator $\langle J_s(z) J_{s'}(w)\rangle$ now arises at second order in $h$. Taking into account normal ordering, for transverse-traceless $h$ one finds
\begin{eqaed}\label{eq:hs_pert_correlator}
    & \langle J_s(z) J_{s'}(0) \rangle_{2} = \int \frac{d^Dk \, d^2x \, d^2y }{\abs{x-y}^{4+\alpha k^2}} \, \norm{R(k)}^2 \, I_s(z) I_{s'}(0)
\end{eqaed}
at second order in $h$, where we have defined
\begin{eqaed}\label{eq:hs_aux_function}
    & I_s(z) \equiv \frac{\alpha^3}{\sqrt{32(2\pi)^D}} \sum_{n=1}^{s-1} (-1)^{n} \, (s-n-1)! (n-1)! \, A_n^s \\
    & \times \left(\frac{1}{(z-x)^{s-n}} - \frac{1}{(z-y)^{s-n}} \right) \left(\frac{1}{(z-x)^{n}} - \frac{1}{(z-y)^{n}} \right) \, .
\end{eqaed}
Once again a lattice regulator is understood, and in order to extract the anomalous dimensions one ought to extract the (quasi-)logarithmic divergences at small $k$. In order to do so, one can rescale $x = z \, u$, $y = z \, v$, so that the overall dependence is $\frac{\abs{z}^{-\alpha k^2}}{z^{s+s'}}$. The holomorphic denominator reflects the tree-level result
\begin{eqaed}\label{eq:hs_tree_correlator}
    \langle J_s(z) J_{s'}(0) \rangle_0 = \frac{\delta_{s s'}}{z^{s+s'}} \, ,
\end{eqaed}
while the numerator carries the leading quasi-logarithmic divergence~\footnote{Let us remark that the UV ``lattice'' cutoff $a$ is always implied in these expressions, since the base of general exponentials is to be dimensionless.} and, indeed, reconstructs the one-loop Ricci flow evaluated at the RG time $t = \log \frac{\abs{z}}{a}$ as in the preceding case. The remainder of \cref{eq:hs_pert_correlator} contains, in general, at most power-like divergences. When the dust settles, the (matrix of) anomalous dimensions $\gamma_{s s'}$ scales, in the IR, according to
\begin{eqaed}\label{eq:hs_anomalous_dims}
    \gamma_{s s'} & \sim \Gamma_{s s'} \int \frac{d^Dk}{(2\pi)^D} \, k^2 \, \norm{R(k)}^2 \, e^{-\alpha k^2 t} \\
    & \sim \Gamma_{s s'} \int \frac{d^Dk}{(2\pi)^D} \, k^2 \, \norm{R(k)}^2 \, e^{-\alpha k^2 \lambda \ell}
\end{eqaed}
up to a (scheme-dependent) constant $\Gamma_{s s'}$, where once again we have factored out the dependence on the RG time for clarity.

As for \cref{eq:scalar_anomalous_dims}, the result in \cref{eq:hs_anomalous_dims} highlights a decay of the anomalous dimensions, and thus of the masses of higher-spin particles, that is precisely exponential in the distance $\ell$, rather than power-like or exponential in a power of $\ell$. Strictly speaking, this is the case for quantized ``momenta'' $k$, as for the case of scalar operators. However, even without compactifying each Fourier mode of the ``graviton vertex'' deformation contributes an exponential decay, albeit the full integral in \cref{eq:hs_anomalous_dims} will in general scale as a negative power of $\ell$ in this case. This result lends further support to the various incarnations of the distance conjecture and to the (S-)duality arguments that we have presented in \cref{sec:bubbles_holography}, remarkably with broken supersymmetry.

\section{Conclusions}\label{sec:conclusions}

The results that we have discussed point to an intriguing mechanism for consistency of string-scale supersymmetry breaking, as well as a novel realization of a number of swampland proposals. The tunneling cascade that we have discussed in \cref{sec:branes_wgc} is closely connected to the weak gravity conjecture~\cite{Antonelli:2019nar, Basile:2021mkd}, and leads to two infinite-distance limits controlled by the flux number $N$. As we have discussed in \cref{sec:bubbles_holography}, in the regime where ten-dimensional EFT is expected to be reliable $N \gg 1$, and the absence of scale separation is reflected by the emergence of a KK tower whose mass scale is exponentially suppressed in the ``discrete-landscape'' distance defined by bubble profiles. The opposite regime, where $N \gtrsim 1$, appears strongly coupled within the EFT description, while the holographic description that we have developed features a decoupled free sector at infinite distance along the dual RG flow.

The free sector restores (super)conformal symmetry, thereby granting stability for the Sugimoto model, and describes a tensionless D-string via conserved single-trace higher-spin currents, whose anomalous dimensions decay in the IR with individual contributions that are exponentially suppressed in the generalized distance that we have introduced in \cref{sec:infinite_distances}. Tantalizingly, emergent supersymmetry appears deeply tied to the proposal of~\cite{Lee:2018urn, Lee:2019xtm, Lee:2019wij} via $\text{Spin}(8)$ triality, and points to a peculiar instance of S-duality. As a result, we are led to speculate that string theory with broken supersymmetry contains the ingredients to remain in the landscape, despite numerous instabilities plaguing its EFT counterpart. Although the present work constitutes but a first step in this direction, and more potential obstacles lurk around the corner, we find the results that we have presented encouraging in this respect. While the approach that we have undertaken can in principle apply to milder supersymmetric settings, it would be interesting to further ground the framework that we have proposed in this paper, in order to sharpen the quantitative grasp of the pressing issues that we have discussed.

\section*{Acknowledgements}

It is a pleasure to thank Jos\'{e} Calder\'{o}n-Infante, Irene Valenzuela, Miguel Montero, Timo Weigand, Connor Behan, Carlo Angelantonj, Federico Carta, Pietro Ferrero, Stefano Lanza, Salvatore Raucci, Giuseppe Bogna, Andrea Luzio, Alessandro Bombini, Davide De Biasio, Simon Pekar, Chrysoula Markou and Evgeny Skvortsov for helpful discussions during the development of this work. I would also like to thank David Tong, Edward Witten and Igor Klebanov for insightful comments about related issues during Strings 2021. I am grateful to Daniel Kl\"{a}wer for useful feedback on the manuscript.

This work was supported by the Fonds de la Recherche Scientifique - FNRS under Grants No.\ F.4503.20 (``HighSpinSymm'') and T.0022.19 (``Fundamental issues in extended gravitational theories'').

\bibliography{bubble_dc}

\end{document}